\documentstyle[preprint,aps,tighten,epsfig,mcite]{revtex}
\begin{document}
\draft
\renewcommand{\thefootnote}{\fnsymbol{footnote}}
\setcounter{footnote}{1}
\title{Floquet-Markov description of the parametrically driven,
      dissipative harmonic quantum oscillator}
\author{Sigmund Kohler, Thomas Dittrich\footnote{Present address:
      Max-Planck-Institut f\"ur Physik komplexer Systeme, Bayreuther
      Stra\ss e 40, Haus 16, D--01187 Dresden, Germany}, and Peter H\"anggi}
\address{Institut f{\"ur} Physik, Universit{\"a}t Augsburg,
      Memminger Stra{\ss}e 6, D--86135 Augsburg, Germany}
\date{\today}
\maketitle
%
\begin{abstract}
Using the parametrically driven harmonic oscillator as a working example,
we study two different Markovian approaches to the quantum dynamics of a
periodically driven system with dissipation. In the simpler approach, the
driving enters the master equation for the reduced density operator only in
the Hamiltonian term. An improved master equation is achieved by treating the
entire driven system within the Floquet formalism and coupling it to the
reservoir as a whole. The different ensuing evolution equations are compared
in various representations, particularly as Fokker-Planck equations for the
Wigner function. On all levels of approximation, these evolution equations
retain the periodicity of the driving, so that their solutions have Floquet
form and represent eigenfunctions of a non-unitary propagator over a single
period of the driving. We discuss
asymptotic states in the long-time limit as well as the conservative and the
high-temperature limits. Numerical results obtained within the different
Markov approximations are compared with the exact path-integral solution.
The application of the improved Floquet-Markov scheme becomes increasingly
important when considering stronger driving and lower temperatures.
\end{abstract}
\section{Introduction}

The dynamics of microscopic systems in strong periodic fields forms a problem
of fundamental significance, with a vast variety of applications in quantum
optics, quantum chemistry, and mesoscopic systems. If the driving field is of
a macroscopic nature, for example, a continuous-wave laser irradiation, it is
appropriate to describe the complete system in a mixed quantum-classical way,
i.e., to give a full quantum-mechanical account of the central system and
its energy loss to ambient degrees of freedom (the electromagnetic vacuum or
weakly coupled internal degrees of freedoms), but to include the field as
a classical external driving force. A solution of the dynamics then requires to
simultaneously eliminate the ambient freedoms and to integrate the equations
of motion with an explicit time dependence. In principle, this can be done
exactly using path-integral techniques. However, even a partially analytical
solution within the path-integral approach is feasible only for the very
simplest systems in the class addressed, in particular, for periodically
driven, damped harmonic oscillators \cite{ZerbeHanggi95},
or for driven dissipative two-level systems
\cite{GrifoniSassettiStockburgerWeiss93,*GrifoniSassettiHanggiWeiss95}.
As soon as nonlinear forces come into
play, the path-integral approach requires to resort to extensive and
sophisticated numerics, such as Monte-Carlo calculations
\cite{Voth93,*MakarovMakri95}, with their own shortcomings.

In most cases of interest, it is more adequate to make as much use as possible
of the methods and approximations that have been developed separately for the
two problems mentioned above, quantum dissipation on the one hand and
periodic driving on the other. Specifically, it is desirable to combine a
Markovian approach to quantum dissipation, leading to a master equation for
the density operator, with the Floquet formalism that allows to treat
time-periodic forces of arbitrary strength and frequency.
While the Floquet formalism amounts essentially to using an optimal
representation and is exact \cite{FainshteinManakovRapoport78},
the simplification brought about by the Markovian description is achieved only
on the expense of accuracy. Here, a subtle technical difficulty lies in the
fact that the truncation of the long-time memory introduced by the bath, and
the inclusion of the driving, do not commute: As pointed out in
Ref.~\cite{GrahamHubner94}, the result of the Markov approximation depends
on whether it is made with respect to the eigenenergy spectrum of the central
system {\it without the driving}, or {\it with} respect to the quasienergy
spectrum obtained from the Floquet solution of the driven system.
In the second case it cannot be treated as a system with
proper eigenstates and eigenenergies.
A Markovian approach based on a quasienergy spectrum has been implemented
in recent work on driven Rydberg atoms
\cite{BlumelBuchleitnerGrahamSirkoSmilanskyWalther91} and driven dissipative
tunneling \cite{DittrichOelschlagelHanggi93,*OelschlagelDittrichHanggi93}.

The purpose of the present paper is to investigate these two Markovian
approaches to damped periodically driven quantum dynamics, with their specific
merits and drawbacks, for a linear system where an exact path-integral
solution is still available: The parametrically driven, damped harmonic
oscillator allows for a very transparent and well-controlled introduction of
the different approximation schemes at hand.  Their quality can here be
reliably checked since in this system, the quasienergy spectrum is
sufficiently different from the unperturbed energy spectrum
\cite{PopovPerelomov70} (this feature is in contrast to the additively driven
harmonic oscillator, where the difference of two quasienergies does not depend
on the driving parameters \cite{PopovPerelomov70}), and a comparison with the
known quantum path-integral solution \cite{ZerbeHanggi95} is possible.

Moreover, by switching to a phase-space representation such as the Wigner
function, it is possible to elucidate the relationship of the quantal results
to the corresponding classical Liouville dynamics. Since this relation is
particularly close in the case of linear systems, this provides an additional
consistency check.  Therefore, the emphasis of this paper is predominantly on
the testing and thorough understanding of the available methods. Their
application to a strongly nonlinear system where analytical path-integral
solutions are far beyond our present capabilities, will be the subject of
forthcoming publications.

Forming a convenient ``laboratory animal'' due to its simplicity and
linearity, the parametrically driven harmonic oscillator still shows
nontrivial behaviour, interesting in its own right. We shall give a brief
review of the model and its classical dynamics in Section II. The central
results of the paper, concerning the applicability and quality of the
alternative Markov approximations, are presented in the course of the
quantization of the system with dissipation, in Section III. Its last
subsection is devoted to a discussion of the asymptotics of the quantal
solutions, such as the conservative and the high-temperature limits. Section
IV contains numerical results for a number of characteristic dynamical
quantities as obtained for the alternative Markovian approaches, and the
comparison to the path-integral solution. A summary of the various
representations and levels of description addressed in the paper, with their
interrelations, is given in Section V. A number of technical issues are
deferred to Appendix \ref{appendix:characteristic}.
Results for an additive time-dependent force in combination with a parametric
periodic driving are summarized in Appendix \ref{appendix:LinearForce}.

%
\section{The model and its classical dynamics}
\label{sect:Classical}
\begin{figure}
\centerline{ \psfig{width=10cm,figure=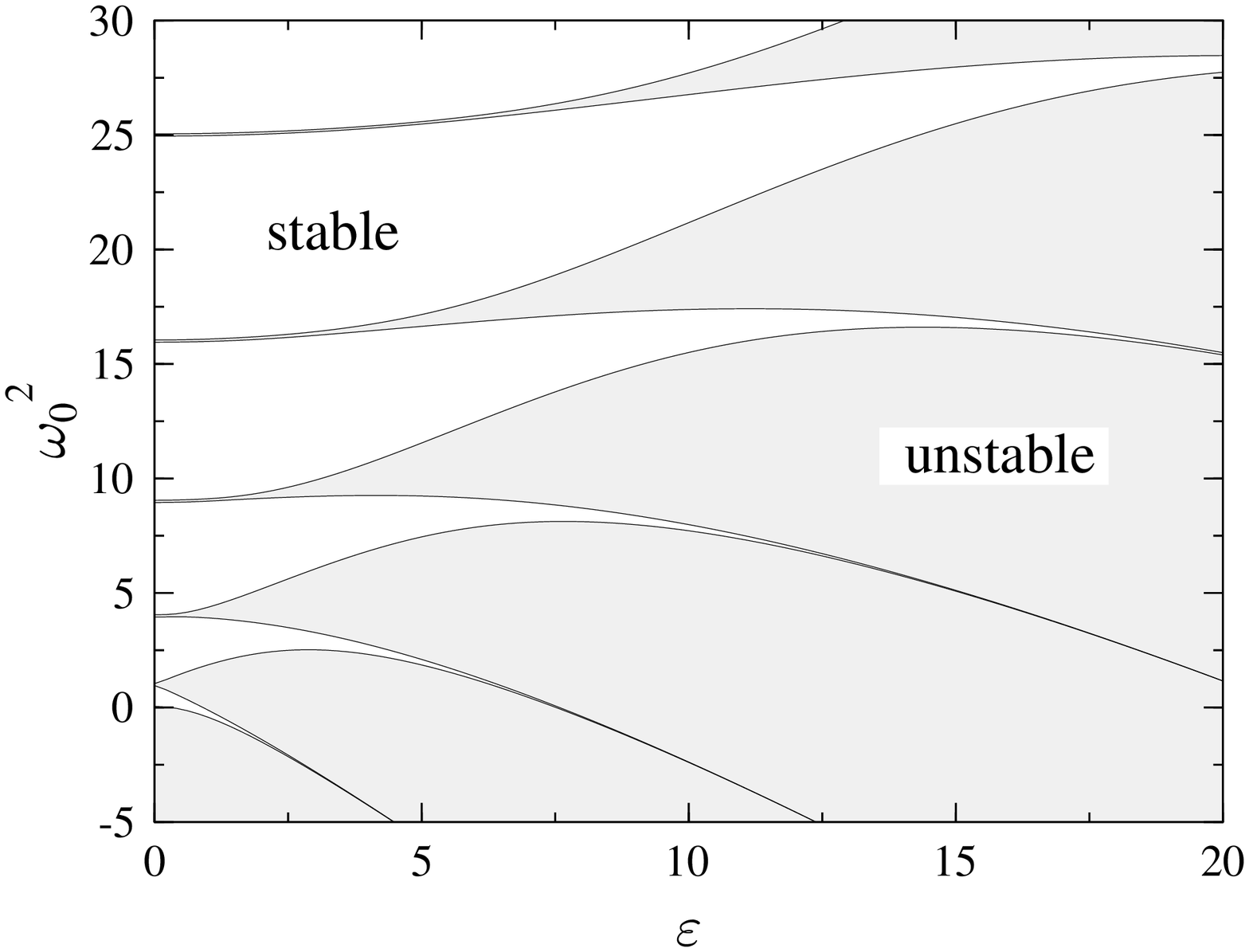} }
\caption{ Stability of equation (\protect\ref{ClassicalX}) with $\gamma=0$
for the case of a Mathieu oscillator.
In the white areas the Floquet index $\mu$ is real, which corresponds to
stable solutions.
In the shaded areas $\mu$ is complex and therefore one of the fundamental
solutions (\protect\ref{FloquetAnsatz}) is unstable.
On the borderlines $\mu$ becomes a multiple of $\Omega$/2.}
\label{fig:stability}
\end{figure}

For a particle with mass $m$ moving in a harmonic potential with 
time-dependent frequency, the Hamiltonian is given by
\begin{equation}
\label{Hamiltonian}
H_{\rm S}(t) = \frac{p^2}{2m} + \frac{1}{2}k(t) x^2,
\end{equation}
where $k(t)$ is a symmetric and periodic function with period $T$.
A special case is the Mathieu oscillator, where
$k(t)=m(\omega_0^2+\varepsilon\cos\Omega t)$ with $\Omega=2\pi/T$. Depending
on its frequency and amplitude, the driving can stabilize or destabilize the
undriven oscillation. Fig.~\ref{fig:stability} shows the zones of stable and
unstable motion, respectively, for the Mathieu oscillator, in the
$\omega_0^2$--$\varepsilon$--plane.
The equation of motion for a classical particle with velocity--proportional
(i.e., Ohmic) dissipation in the potential given in (\ref{Hamiltonian}) reads
\begin{equation}
\label{ClassicalX}
\ddot x + \gamma \dot x + \frac{1}{m}k(t) x = 0.
\end{equation}
By substituting $x=y\exp(-\gamma t/2)$, we can formally remove the
damping to get an undamped equation with a modified potential
\begin{equation}
\label{ClassicalY}
\ddot y + \left( k(t)/m - \gamma^2/4 \right) y = 0.
\end{equation}
Already here, on the level of the classical equations of motion, we can apply
the Floquet theorem for second-order differential equations with time-periodic
coefficients. It asserts \cite{MagnusWinkler79,*McLachlan64} that
Eq.~(\ref{ClassicalY}) has two solutions of the form
\begin{equation}
\xi_1(t) = {\rm e}^{{\rm i}\mu t}\varphi(t), \quad
\xi_2(t)=\xi_1(-t), \quad
\varphi(t+T) = \varphi(t).
\label{FloquetAnsatz}
\end{equation}
The solution $\xi_2(t)$ is related to $\xi_1(t)$ by the time-inversion
symmetry inherent in (\ref{ClassicalY}). Being periodic in time, the
classical Floquet function $\varphi(t)$ can be represented as a
Fourier series
\begin{equation}
\label{FourierAnsatz}
\varphi(t) = \sum_{n=-\infty}^\infty c_n {\rm e}^{{\rm i}n\Omega t}.
\end{equation}
The Floquet index $\mu$ depends on the shape of the driving $k(t)$ and is
defined only ${\,\rm mod\,} \Omega$. There exist driving functions for which
$\mu$ is complex so that one of the solutions $\xi_i(t)$ becomes unstable
(cf.\ Fig.~\ref{fig:stability}).
In stable regimes $\mu$ is real.
On the border between a stable and an unstable regime, $\mu$ becomes a multiple
of $\Omega/2$ and the solutions $\xi_1(t)$ and $\xi_2(t)$ are not linearly
independent.  For given $k(t)$,
the $\xi_i(t)$ still depend on the damping $\gamma$. We denote the limit
$\gamma\to 0$ of the functions $\xi_i(t)$ by $\xi^{0}_i(t)$.

The normalization of the $c_n$ is chosen such that the Wronskian $\cal W$,
which is a constant of motion, is given by
\begin{equation}
\label{Wronskii}
{\cal W} = \dot\xi_1(t)\xi_2(t) - \xi_1(t)\dot\xi_2(t) = 2{\rm i},
\end{equation}
resulting in the sum rule
\begin{equation}
\label{SumRule}
\sum_{n=-\infty}^\infty c_n^2 (\mu+n\Omega) = 1 .
\end{equation}

Returning to the original $x$--coordinate, we find that the fundamental
solutions of (\ref{ClassicalX}) read
\begin{equation}
f_i(t) = {\rm e}^{-\gamma t/2}\xi_i(t), \qquad i=1,2.
\end{equation}

For constant frequency of the oscillator, $k(t)={\rm const}=m\omega_0^2$, the
Floquet index and the periodic function become $\mu =
(\omega_0^2-\gamma^2/4)^{1/2}$ and $\varphi(t) =
(\omega_0^2-\gamma^2/4)^{-1/2}$, respectively, which reproduces the results
for a damped harmonic oscillator without driving.

The Green function for Eq.~(\ref{ClassicalX}) is constructed using
Eqs.~(\ref{FourierAnsatz}) and (\ref{Wronskii}),
\begin{eqnarray}
\label{ClassicalGreenFunction}
G(t,t') &=&
{\rm e}^{-\gamma(t-t')/2}
\left[ \xi_1(t)\xi_2(t')-\xi_2(t)\xi_1(t') \right]/2{\rm i} \\
\label{ClassicalGreenFunctionFourier}
&=&
{\rm e}^{-\gamma(t-t')/2}\sum_{n,n'}c_n c_{n'}
\sin\left[ \mu(t-t')+\Omega(nt-n't')\right] .
\end{eqnarray}
In terms of this function, the solution of (\ref{ClassicalX}) with initial
conditions $x(t_0)=x_0$ and $p(t_0)=p_0$, reads
\begin{equation}
\label{ClassicalSolution}
x(t,t_0) = -x_0\frac{\partial G(t,t_0)}{\partial t_0} + \frac{p_0}{m} G(t,t_0).
\end{equation}
Since the potential breaks continuous time-translational invariance, this
solution depends explicitly on the initial time $t_0$.

\section{The dissipative quantum system}
To achieve a microscopic model of dissipation, we couple the system
(\ref{Hamiltonian}) bilinearly to a bath of non-interacting harmonic
oscillators \cite{Zwanzig73}. 
The total Hamiltonian of system and bath is then given by
\begin{equation}
\hat H(t) = \hat H_{\rm S}(t) + \hat H_{\rm SB} + \hat H_{\rm B},
\end{equation}
where
\begin{equation}
\label{HamiltonianB}
\hat H_{\rm B} = \sum_{\nu=1}^{N} \left( \frac{\hat p_\nu^2}{2m_\nu}
     +\frac{m_\nu}{2}\omega_\nu^2 \hat x_\nu^2 \right)
\end{equation}
is the Hamiltonian of $N$ oscillators with masses $m_\nu$,
frequencies $\omega_\nu$, momenta $\hat p_\nu$, and coordinates $\hat x_\nu$.
The bath interacts with the system via
\begin{equation}
\label{HamiltonianSB}
\label{InteractionHamiltonian}
\hat H_{\rm SB} = -\hat x \sum_{\nu=1}^{N} g_\nu \hat x_\nu
         +\hat x^2\sum_{\nu=1}^{N}\frac{g_\nu^2}{2m_\nu\omega_\nu^2} ,
\end{equation}
which couples the system to each bath oscillator $\nu$ with a strength
$g_\nu$.  The second term in Eq.~(\ref{InteractionHamiltonian}) serves to
cancel a shift of the potential minimum due to the coupling
\cite{Zwanzig73,HanggiTalknerBorkovec90,*Weiss93}.
The bath is fully characterized by the spectral density of the coupling
energy,
\begin{equation}
\label{SpectralDensity}
I(\omega) = \pi\sum_{\nu=1}^N \frac{g_\nu^2}{2m_\nu\omega_\nu}
\delta(\omega-\omega_\nu).
\end{equation}

We choose an initial condition of the Feynman-Vernon type, i.e.,
at $t=t_0$ the bath is in thermal equilibrium and uncorrelated to the
system, i.e.
\begin{equation}
\label{FVinitial}
\rho(t_0) = \rho_{\rm S}(t_0)\otimes\rho_{\rm B,eq},
\end{equation}
where $\rho_{\rm B,eq}=\exp(-\hat H_{\rm B}/k_{\rm B}T)/{\rm tr}_{\rm B}
\exp(-\hat H_{\rm B}/k_{\rm B}T)$
is the canonical ensemble of the bath and $k_{\rm B}T$ Boltzmann's constant
times temperature.

\subsection{Interaction picture and perturbation theory}
Due to the bilinearity of the system-bath coupling, one can always eliminate
the bath variables to get an exact, closed integro-differential equation for
the reduced density matrix $\rho_{\rm S}={\rm tr}_{\rm B}\rho$, which
describes the dynamics of the central system, subject to dissipation
\cite{Haake73,AlickiLendi87,Louisell73}.
In most cases, however, this equation cannot be solved exactly.
In the limit of weak coupling,
\begin{eqnarray}
\gamma & \ll & k_{\rm B}T/\hbar,\\
\gamma & \ll & \Delta_{\alpha\beta},
\end{eqnarray}
it is possible to truncate the time-dependent perturbation expansion in the
system-bath interaction after the second-order term. The quantity
$\gamma$ denotes the effective damping of the dissipative system,
and $\Delta_{\alpha\beta}$ are the transition frequencies of the central
system (see, e.g.~Eq.~(\ref{TransFreq}), below).
The autocorrelations of the bath decay on a time scale $\hbar/k_{\rm B}T$, and
thus in this limit instantaneously on the time scale $1/\gamma$ of the system
correlations.

With the initial preparation (\ref{FVinitial}), the equation of
motion for the reduced density matrix in this approximation is given by
\cite{Haake73,AlickiLendi87,Louisell73}
\begin{eqnarray}
\nonumber
\dot \rho_{\rm S}(t)
&=& -\frac{{\rm i}}{\hbar}\left[ \hat H_{\rm S}(t),\rho_{\rm S}(t) \right]
   -\frac{{\rm i}}{\hbar} {\rm tr}_{\rm B} \left[ \hat H_{\rm SB},
    \rho_{\rm S}(t) \right] \\
&& - \frac{1}{\hbar^2} \int_0^\infty {\rm d}\tau\, {\rm tr}_{\rm B}
   \left[\hat H_{\rm SB},\left[\tilde H_{\rm SB}(t-\tau,t),
   \rho_{\rm B,eq}\otimes\rho_{\rm S}(t)
   \right]\right] .
\label{MasterEquationGeneral}
\end{eqnarray}
The tilde denotes the interaction picture defined by
\begin{eqnarray}
\tilde{\cal O}(t,t') &=& U_0^\dagger(t,t')\hat {\cal O}U_0(t,t'), \\
U_0(t,t')&=& {\cal T}\exp\left(-\frac{{\rm i}}{\hbar}
	\int_{t'}^t{\rm d}t''(\hat H_{\rm S}(t'')+\hat H_{\rm B}) \right) ,
\end{eqnarray}
where ${\cal T}$ is Wick's time-ordering operator.

For $\hat H_{\rm S}$ and $\hat H_{\rm SB}$ as in Eqs.~(\ref{HamiltonianB}),
(\ref{HamiltonianSB}), we find the master equation
\begin{eqnarray}
\nonumber
\dot \rho_{\rm S}(t)
&=& -\frac{{\rm i}}{\hbar}\left[ \hat H_{\rm S}(t),\rho_{\rm S}(t) \right] \\
\nonumber
&& -\frac{1}{\hbar^2} \sum_{\nu=1}^N g_\nu^2
   \int_0^\infty {\rm d}\tau
   \Big\{ S_\nu(\tau)\big[\hat x,\big[\tilde x(t-\tau,t),
  \rho_{\rm S}(t)\big]\big] \\
&& \qquad\qquad\qquad
   + {\rm i} A_\nu(\tau)
   \big[\hat x,\big[\tilde x(t-\tau,t),\rho_{\rm S}(t)\big]_+\big]\Big\},
\label{MasterEquation}
\end{eqnarray}
with $[A,B]_+=AB+BA$ and
\begin{eqnarray}
\label{BathCorrelationS}
S_\nu(t) &=& \frac{\hbar}{2m_\nu\omega_\nu}
\coth\left(\frac{\hbar\omega_\nu}{2k_{\rm B}T}\right)
\cos\omega_\nu t,\\
\label{BathCorrelationA}
A_\nu(t) &=& -\frac{\hbar}{2m_\nu\omega_\nu} \sin\omega_\nu t,
\end{eqnarray}
the symmetrically ordered and antisymmetrically ordered, respectively,
correlation functions of the bath oscillator $\nu$.

\subsection{Markov approximation with respect to the unperturbed spectrum}
\label{sect:SimpleMarkov}
So far, we have followed the standard approach to dissipative quantum dynamics
in the weak coupling limit \cite{Haake73,AlickiLendi87,Louisell73}.
In the following subsections, we shall contrast a
simpler Markov approximation based on the unperturbed spectrum, with a more
sophisticated approach that accounts for the modification of the spectrum due
to the driving.

\subsubsection{Master equation}
\label{sect:MasterEq}
In the following, we restrict ourselves to an ohmic bath,
\begin{equation}
I(\omega)=m\gamma\omega,
\label{Ohmic}
\end{equation}
fixing the relation between the macroscopic damping constant $\gamma$ and the
microscopic coupling constants $g_n$ introduced in
Eq.~(\ref{InteractionHamiltonian}).  By imposing a Drude cutoff
$I(\omega)\to I(\omega)/(1-{\rm i}\omega/\omega_{\rm D})$
with $\omega_{\rm D} \gg \omega_0,\Omega$, divergent integrals are avoided.

In a crudest approximation, the time dependence of the system
Hamiltonian is neglected in the derivation of the master equation,
i.e., the incoherent terms in the master equation are calculated replacing
$\hat H_{\rm S}(t)$ by
$\bar H_{\rm S}=(1/T)\int_0^T{\rm d}t\,\hat H_{\rm S}(t)$,
i.e. the Hamiltonian with zero driving amplitude.
The position operator in the interaction picture is then given by
\begin{equation}
\tilde x(t,t')=\hat x\cos\omega_0(t-t') +
\frac{\hat p}{m\omega_0}\sin\omega_0(t-t') .
\end{equation}
Since the information on the phase of the driving is lost, it depends only on
the difference $t-t'$ of its arguments.

Inserting this operator and the correlation functions
(\ref{BathCorrelationS}), (\ref{BathCorrelationA})
into Eq.~(\ref{MasterEquation}), leads to the master equation
\begin{eqnarray}
\nonumber
\dot\rho_{\rm S} &=&
 -\frac{{\rm i}}{\hbar}\big[\hat H_{\rm S}(t),\rho_{\rm S}\big]
 -\frac{{\rm i}}{2\hbar}\bar\gamma\ [\hat x,[\hat p,\rho_{\rm S}]_+] \\
&&
 -\frac{\gamma}{\hbar^2}D_{pp} [\hat x,[\hat x,\rho_{\rm S}]]
 +\frac{\gamma}{\hbar^2}D_{xp} [\hat x,[\hat p,\rho_{\rm S}]] .
\label{MasterEquationSimple}
\end{eqnarray}
The right-hand side of this equation depends on time only through its first,
the Hamiltonian, term and therefore retains the periodicity of the system
Hamiltonian $\hat H_{\rm S}(t)$.

This form of the master equation does not produce a positive
semidefinite diffusion matrix. It consequently does not exhibit Lindblad form
\cite{AlickiLendi87,Lindblad76,Talkner86,Diosi93,*Diosi93b}.  The positivity
of $\rho_{\rm S}$ is thus not guaranteed for all elements of the function
space of density operators. The Markovian approximation implies that quantum
effects on a length scale $l<\lambda_{\rm dB}=\hbar/\sqrt{4mk_{\rm B}T}$
(non-Markov effects) cannot be described selfconsistently
\cite{Diosi93,*Diosi93b,Pechukas90,Ambegaokar91}.  Note also that within a
Markov approximation, the master equation is periodic with the driving period
$T=2\pi/\Omega$ (Floquet form).  This is in contrast to the non-Markovian
exact master equation \cite{ZerbeHanggi95}. In this latter case, the effective
master equation has the structure of (\ref{MasterEquationSimple}) with
time-dependent coefficients $D_{xp}$ and $D_{pp}$ that depend also in a
non-periodic way on the time elapsed since the preparation at $t_0$. In Wigner
representation, this corresponds to a time-dependent diffusion coefficient
(see below).

The coefficients $\bar\gamma$ and $D_{pp}$ can be evaluated
straightforwardly \cite{OppenheimRomerorochin87} to give
\begin{eqnarray}
\bar\gamma &=& \gamma , \\
\label{DppSimple}
D_{pp} &=&
\frac{1}{2}m\hbar\omega_0 \coth \frac{\hbar\omega_0}{2k_{\rm B}T} .
\end{eqnarray}

The evaluation of the cross-diffusion $D_{xp}$ is more complex.
Because we did not find it in the literature, we give the outline of
its derivation.
The logarithmic divergence of $D_{xp}$ is regularized by the Drude cutoff
to obtain
\begin{equation}
\label{Dxp}
D_{xp} = -\frac{\hbar}{2\pi}{\rm P}\int_{-\infty}^\infty {\rm d}\omega
\coth\left(\frac{\hbar\omega}{2k_{\rm B}T}\right)
\frac{\omega}{\omega^2-\omega_0^2}
\; \frac{{\rm i}\omega_{\rm D}}{\omega+{\rm i}\omega_{\rm D}} ,
\end{equation}
where P denotes Cauchy's principal part.
The integral in Eq.~(\ref{Dxp}) is solved by contour integration in the upper
half plane.
Expressing the resulting sums by the psi function
$\psi(x)={\rm d}\ln\Gamma(x)/{\rm d}x$ \cite{GradshteynRyzhik}
and neglecting terms of the order $\omega_0/\omega_{\rm D}$, we obtain
\begin{equation}
\label{DxpHighCutoff}
D_{xp} = -\frac{\hbar}{\pi} \left[
   \psi\left(1+{\hbar\omega_{\rm D} \over 2\pi k_{\rm B}T}\right)+ C \right],
\end{equation}
where $C$ is the Euler constant.

Interestingly enough, $m\gamma D_{xp}$ coincides with the Drude
regularized divergent
part of the stationary momentum variance of a dissipative harmonic oscillator
\cite{GrabertWeissTalkner84,*RiseboroughHanggiWeiss85,*GrabertSchrammIngold88}.

It must be stressed that the dissipative terms in the master equation
(\ref{MasterEquationSimple}) are {\it independent of the driving\/}.
This manifestly reflects that the time dependence of $H_{\rm S}(t)$ has not
been taken into account in the incoherent terms of the master equation.

\subsubsection{Wigner representation and Fokker-Planck equation}
In order to achieve a description close to the classical phase-space dynamics,
we discuss the time evolution of the density operator in Wigner representation.
It is defined by \cite{Wigner32,*HilleryOConnellScullyWigner84}
\begin{equation}
W(x,p,t) = \frac{1}{\pi\hbar}\int_{-\infty}^\infty {\rm d}x'
{\rm e}^{{2 \rm i}px'/\hbar}
\langle x-x' | \rho_{\rm S}(t) | x+x' \rangle .
\end{equation}
The moments of the Wigner function are the symmetrically-ordered
expectation values of the density operator.

Applying this transformation to the master equation
(\ref{MasterEquation}), we obtain a c-number equation of motion,
\begin{equation}
\label{FokkerPlanck}
\partial_t W(x,p,t) = L(t) W(x,p,t) ,
\end{equation}
with the differential operator
\begin{equation}
\label{FokkerPlanckOperator}
L(t)=
   -{1\over m}p\partial_x + \gamma\partial_p p + k(t)x\partial_p
   +\gamma D_{pp}\partial_p^2
   +\gamma D_{xp}\partial_x \partial_p .
\end{equation}
Equation (\ref{FokkerPlanckOperator}) has the structure of an effective
Fokker-Planck operator. However, for $D_{xp}\neq 0$, the diffusion
matrix is not positive semidefinite; correspondingly (\ref{FokkerPlanck})
has no equivalent Langevin representation.

As is the case for the master equation from which it has been derived, the
coefficients of the Fokker-Planck operator retain the it periodicity of the
driving, so that Eq.~(\ref{FokkerPlanck}) has solutions of Floquet form. This
fact will be exploited in the following subsection to construct the solutions.

\subsubsection{Wigner-Floquet solutions}
The Fokker-Planck equation for the density operator in Wigner representation,
Eq.~(\ref{FokkerPlanck}) with Eq.~(\ref{FokkerPlanckOperator}), offers
the opportunity to make full use of the well-known and intuitive results for
the corresponding classical stochastic system. In particular, a solution of
the Fokker-Planck equation can be obtained directly by solving the equivalent
Langevin equation \cite{Risken84,HanggiThomas82}, or by using the formula for
the conditional
probability of a Gauss process \cite{HanggiThomas82}. In the present case,
however, the fact that the diffusion matrix of (\ref{FokkerPlanckOperator}) is
not positive semidefinite requires to take a different route.

Since Eq.~(\ref{FokkerPlanck}) with Eq.~(\ref{FokkerPlanckOperator})
represents a differential equation with time-periodic coefficients, it
complies with the conditions of the Floquet theorem. Consequently, there
exists a complete set of solutions of the form
\begin{equation}
W_\alpha(x,p,t) = {\rm e}^{\mu_\alpha t} u_\alpha(x,p,t), \quad
u_\alpha(x,p,t)=u_\alpha(x,p,t+T),
\end{equation}
henceforth referred to as {\it Wigner-Floquet functions\/}. 

We construct a solution for (\ref{FokkerPlanck}) of this form with
$\mu_{00}=0$ by the method of characteristics \cite{Kamke79b}, cf.\ the
Appendix \ref{appendix:characteristic}.
In the limit $t_0\to -\infty$, the terms in the first line of
(\ref{S(X,P,t)}), which contain the initial condition, vanish and we obtain the
asymptotic solution
\begin{equation}
W_{00}(x,p,t) = 
\frac{1}{2\pi}
\left| \begin{array}{cc}
   \sigma_{xx}(t) & \sigma_{xp}(t) \\
   \sigma_{xp}(t) & \sigma_{pp}(t)
\end{array}\right|^{-1/2}
\exp\left\{-\frac{1}{2}
\left( \begin{array}{c} x \\ p \end{array}\right)
\left( \begin{array}{cc}
   \sigma_{xx}(t) & \sigma_{xp}(t) \\
   \sigma_{xp}(t) & \sigma_{pp}(t)
\end{array}\right)^{-1}
\left( \begin{array}{c} x \\ p \end{array}\right)
\right\}
\end{equation}
with the variances
\begin{eqnarray}
\label{SigmaXX}
\sigma_{xx}(t) &=&
	\frac{2\gamma D_{pp}}{m^2}\int_{-\infty}^t {\rm d}t'
	\left[ G(t,t') \right]^2 , \\
\label{SigmaXP}
\sigma_{xp}(t) &=&
	\frac{2\gamma D_{pp}}{m}\int_{-\infty}^t {\rm d}t'
	G(t,t') \frac{\partial}{{\partial}t}G(t,t') , \\
\label{SigmaPP}
\sigma_{pp}(t) &=& - m\gamma D_{xp}
	+ 2\gamma D_{pp}\int_{-\infty}^t {\rm d}t'
	\left[ \frac{\partial}{{\partial}t}G(t,t')\right]^2 .
\end{eqnarray}
Note that in (\ref{SigmaXX})--(\ref{SigmaPP}) the difference in using 
$D_{pp}$ and $D=D_{pp}+\gamma D_{xp}$
(see (\ref{Deffective}) in Appendix \ref{appendix:characteristic})
is meaningless, since it is a correction of order $\gamma$.
By inserting for $G(t,t')$ the Fourier representation
(\ref{ClassicalGreenFunctionFourier}), one finds that the variances are
asymptotically time-periodic.

Starting from $W_{00}$, we construct further Wigner-Floquet functions:
By solving the characteristic equations
(see Appendix \ref{appendix:characteristic}),
we find the two time-dependent differential operators
\begin{eqnarray}
Q_{1+}(t) &=& f_1(t)\partial_x + m \dot f_1(t)\partial_p , \\
Q_{2+}(t) &=& f_2(t)\partial_x + m \dot f_2(t)\partial_p .
\end{eqnarray}
They have the properties
\begin{equation}
\label{VRplus}
\left[ L(t)-\partial_t,Q_{1+}(t)\right] =
\left[ L(t)-\partial_t,Q_{2+}(t)\right] = 0
\end{equation}
and
\begin{eqnarray}
\label{FloquetQ1}
Q_{1+}(t+T) &=& {\rm e}^{(-\gamma/2+{\rm i}\mu)T} Q_{1+}(t) , \\
\label{FloquetQ2}
Q_{2+}(t+T) &=& {\rm e}^{(-\gamma/2-{\rm i}\mu)T} Q_{2+}(t) .
\end{eqnarray}
Taking the commutation relation (\ref{VRplus}) into account, the functions
\begin{equation}
\label{Eigenfunctions}
W_{nn'}(x,p,t) = Q_{1+}^n(t) Q_{2+}^{n'}(t) W_{00}(x,p,t), \quad
n,n' = 0,1,2,\ldots
\end{equation}
also solve Eq.~(\ref{FokkerPlanck}).

Due to Eqs.~(\ref{FloquetQ1}), (\ref{FloquetQ2}), they are of Floquet
structure with the Floquet spectrum
\begin{equation}
\mu_{nn'} = n(-\gamma/2 + {\rm i}\mu) + n'(-\gamma/2 - {\rm i}\mu) .
\end{equation}
This spectrum is independent of the diffusion constants, as expected for an
operator of type (\ref{FokkerPlanckOperator})
\cite{HwaliszJungHanggiTalknerSchimanskygeier89}, and therefore is the
same as in the case of classical parametrically driven Brownian oscillator
\cite{ZerbeJungHanggi94}.

The expression for the eigenfunctions in the high-temperature limit of
the (undriven) classical Brownian harmonic oscillator in
Refs.~\cite{HwaliszJungHanggiTalknerSchimanskygeier89,Titulaer78}
is also of the structure (\ref{Eigenfunctions}).
We can recover this solution by inserting the classical diffusion constant
$mk_{\rm B}T$ and the undriven limit $\varepsilon\to 0$ for the classical
solution, given in Sect.~\ref{sect:Classical}.

\subsection{Markov approximation with respect to the quasienergy spectrum}
The master equation (\ref{MasterEquationSimple}) can be improved
by including the time-dependent term in the system Hamiltonian
(\ref{Hamiltonian}) before a Markov approximation is introduced, to account
for the change in the quasienergy spectrum due to the driving.

\subsubsection{Floquet theory and quasienergy spectrum}
For a Schr\"odinger equation with time-periodic system Hamiltonian such as
(\ref{Hamiltonian}), the Floquet theorem \cite{FainshteinManakovRapoport78}
asserts that there exists a complete set of solutions of the form
\begin{equation}
\label{SchrodingerFloquetStates}
|\psi_\alpha(t)\rangle = {\rm e}^{-{\rm i}\mu_\alpha t}|\phi_\alpha(t)\rangle,
\quad |\phi_\alpha(t+T)\rangle=|\phi_\alpha(t)\rangle
\end{equation}
The quasienergy $\mu_\alpha$ plays the role of a phase and therefore is only
defined ${\,\rm mod\,} \Omega$, cf.\ Ref.~\cite{FainshteinManakovRapoport78}.
We shall use the basis $\{|\psi_\alpha(t)\rangle\}$ as an optimal
representation to decompose states and operators.

For the parametrically driven harmonic oscillator (\ref{Hamiltonian}),
the Floquet solutions for the Schr\"odinger equation are derived in the
literature in various ways
\cite{PopovPerelomov69,LewisRiesenfeld69,Brown91,SchradeMankoSchleichGlauber95}.
We skip the derivation and merely present the result,
\begin{equation}
\label{FloquetSchrodinger}
\psi_\alpha(x,t)=
\left( \frac{\sqrt{m/\pi\hbar}}{2^\alpha n!\xi^{0}_1(t)} \right)^{1/2}
\left(\frac{\xi^{0}_1(t)}{\xi^{0}_2(t)}\right)^{\alpha/2}
H_\alpha\left(x\sqrt{m/\hbar \xi^{0}_1(t)\xi^{0}_2(t)}\right)
\exp\left({\rm i} {\dot\xi^{0}_1(t)x^2}/2{\xi^{0}_1(t)}\right),
\end{equation}
for the Floquet solutions in the stable regime, where $H_\alpha$ is the
$\alpha$-th Hermite polynomial, $\alpha=0,1,2,\ldots$.
The Floquet index for this solution is $\mu_\alpha=\mu(\alpha+1/2)$.
This gives the quasienergy spectrum
\begin{equation}
\label{SchrodingerFloquetSpectrum}
\mu_{\alpha,k}=(\alpha+1/2)\mu^{0} + k\Omega, \quad
k = 0,\pm 1,\pm 2,\ldots.
\end{equation}
Note that (\ref{FloquetSchrodinger}) are solutions only in the stable regime.
Consequently $\mu$ is real, cf.~Sect.~\ref{sect:Classical}.

In analogy to the annihilation and creation operators for the undriven
harmonic oscillator, one can define operators $\hat\Gamma$ and
$\hat\Gamma^\dagger$ which act as shift operators for the Floquet states, i.e.
\begin{eqnarray}
\label{Annihilation}
\hat\Gamma(t)|\psi_\alpha(t)\rangle 
&=& \sqrt{\alpha}\, |\psi_{\alpha-1}(t)\rangle ,\\
\label{Creation}
\hat\Gamma^\dagger(t)|\psi_\alpha(t)\rangle
&=&\sqrt{\alpha+1}\, |\psi_{\alpha+1}(t)\rangle.
\end{eqnarray}
For a parametrically driven harmonic oscillator, $\hat\Gamma(t)$
can be expressed in terms of position and momentum operator as
\cite{LewisRiesenfeld69,Brown91}
\begin{equation}
\label{ShiftOperator}
\hat\Gamma(t) =
   \frac{1}{2\rm i}\left( \hat x \sqrt{\frac{2m}{\hbar}}\dot\xi_1^{0}(t)
 - \hat p \sqrt{\frac{2}{m\hbar}}\xi_1^{0}(t) \right).
\end{equation}
The relations (\ref{Annihilation}) and (\ref{Creation}) can be proven
by inserting the Floquet solutions (\ref{FloquetSchrodinger}) and using
the recursion relations for Hermite polynomials
\cite{GradshteynRyzhik}.

The matrix element $X_{\alpha\beta}(t)$ of the position operator $x$ with
the states $|\psi_\alpha(t)\rangle$, which we shall need later, reads
\begin{eqnarray}
X_{\alpha\beta}(t)
&=& {\rm e}^{{\rm i}(\mu_\alpha-\mu_\beta)t}
	\langle\phi_\alpha(t)|x|\phi_\beta(t)\rangle \\
\label{Xeval2}
&=& \sum_k {\rm e}^{{\rm i}\Delta_{\alpha\beta k}t}
	X_{\alpha\beta k} ,\\
\label{Xeval3}
\label{XFourier}
X_{\alpha\beta k}
&=& \frac{1}{T}\int_0^T{\rm d}t\, {\rm e}^{-{\rm i}k\Omega t}
\langle\phi_\alpha(t)|x|\phi_\beta(t)\rangle,
\end{eqnarray}
with the transition frequencies
\begin{equation}
\label{TransFreq}
\Delta_{\alpha\beta k}=\mu_\alpha-\mu_\beta + k\Omega.
\end{equation}
For Eqs.~(\ref{Xeval2}) and (\ref{Xeval3}), the periodicity of the Floquet
states $|\phi_\alpha(t)\rangle$ has been used. The Fourier components
$X_{\alpha\beta k}$ are preferably evaluated in the spatial representation,
\begin{eqnarray}
X_{\alpha\beta}(t) 
&=& \int_{-\infty}^\infty {\rm d}x\,\psi_\alpha(x,t)\,x\,\psi_\beta(x,t) \\
&=& \sqrt{\frac{\hbar}{2m}}
	\left(\sqrt{\beta}\xi^{0}_2(t)\delta_{\alpha,\beta-1}+
	\sqrt{\alpha}\xi^{0}_1(t)\delta_{\alpha,\beta+1}\right) ,
\end{eqnarray}
by inserting the Fourier expansion (\ref{FourierAnsatz}) for $\xi^{0}_i(t)$,
to give
\begin{equation}
\label{XFloquetFourier}
X_{\alpha\beta k} = \sqrt{\frac{\hbar}{2m}}
\left(\sqrt{\beta}\,c_{-k}\delta_{\alpha,\beta-1}+
\sqrt{\alpha}\,c_k\delta_{\alpha,\beta+1}\right).
\end{equation}

\subsubsection{Improved master equation}

We start anew from the full master equation in the weak-coupling limit,
\begin{eqnarray}
\nonumber
\dot\rho
&=&
-\frac{{\rm i}}{\hbar}\big[ \hat H_{\rm S}(t),\rho \big] \\
&&
+\frac{1}{\pi\hbar}\int_{-\infty}^\infty {\rm d}\omega\, I(\omega)
n_{\rm th}(\omega)\int_0^\infty {\rm d}\tau\,{\rm e}^{{\rm i}\omega\tau}
\big[\tilde x(t-\tau,t)\rho,\hat x \big] + \text{h.c.}\;.
\end{eqnarray}
Here, h.c. denotes the hermitian conjugate of the dissipative part and
\begin{equation}
n_{\rm th}(\omega)
= \left({\rm e}^{\hbar\omega/k_{\rm B}T}-1\right)^{-1}
= -n_{\rm th}(-\omega)-1
\end{equation}
gives the thermal occupation of the bath oscillator with frequency $\omega$.
To achieve a more compact notation, we have required that $I(-\omega) =
-I(\omega)$, which for an Ohmic bath, cf.\ Eq.~(\ref{Ohmic}), is just the
analytic continuation.

The fact that the Floquet states $|\psi_\alpha(t)\rangle$ of the undamped
central system, Eq.~(\ref{SchrodingerFloquetStates}), solve the Schr\"odinger
equation, allows for a substantial formal simplification of the master
equation: With the density operator being represented in this basis,
\begin{equation}
\rho_{\alpha\beta}(t)
= \langle\psi_\alpha(t)| \rho(t) |\psi_\beta(t)\rangle,
\end{equation}
the master equation takes the form
\begin{eqnarray}
\nonumber
\dot\rho_{\alpha\beta} =
\frac{1}{\pi\hbar} && \int_{-\infty}^\infty {\rm d}\omega\,I(\omega)
n_{\rm th}(\omega) \\
\times && \int_0^\infty {\rm d}\tau\,{\rm e}^{{\rm i}\omega\tau}
\sum_{\alpha'\beta'}\left\{
X_{\alpha\alpha'}(t-\tau)\rho_{\alpha'\beta'}X_{\beta\beta'}^\ast(t)
-X_{\alpha'\alpha}^\ast(t) X_{\alpha'\beta'}(t-\tau)\rho_{\beta'\beta}
\right\} +\text{h.c.}
\end{eqnarray}

Inserting (\ref{XFourier}) and (\ref{XFloquetFourier}) and using the
identity $\int_0^\infty {\rm d}\tau\,
{\rm e}^{{\rm i}\omega\tau} =  \pi\delta(\omega) + {\rm P}({\rm i}/\omega)$,
we arrive at the explicit equation of motion
\begin{eqnarray}
\nonumber
\dot\rho_{\alpha\beta}
&=&
\frac{1}{\hbar} \sum_{\alpha'\beta'}\sum_{kk'} \Big\{
-I(\Delta_{\alpha'\beta'k'})n_{\rm th}(\Delta_{\alpha'\beta'k'})
{\rm e}^{{\rm i}(\Delta_{\alpha'\beta'k'}-\Delta_{\alpha\alpha'k})t}
X_{\alpha'\alpha k}^\ast X_{\alpha'\beta'k'} \rho_{\beta'\beta}\\
&&
+I(\Delta_{\alpha\alpha'k})n_{\rm th}(\Delta_{\alpha\alpha'k})
{\rm e}^{{\rm i}(\Delta_{\alpha\alpha'k}-\Delta_{\beta\beta'k'})t}
X_{\alpha'\alpha k} \rho_{\alpha'\beta'} X_{\beta\beta'k'}^\ast \Big\}
+\text{h.c.} .
\label{Improved MasterEquation}
\end{eqnarray}
The quasienergies of the undamped central system appear in Eq.~(\ref{Improved
MasterEquation}) by way of the $\Delta_{\alpha\beta k}$.  Since these
frequencies contain only differences of quasienergies, they have a direct
physical significance as transition frequencies and so may be used as
arguments of $I(\omega)$ and $n_{\rm th}(\omega)$. This is not the case for
the quasienergies themselves, due to their Brillouin-zone-like ambiguity, cf.\
Eq.~(\ref{SchrodingerFloquetSpectrum}). Shifts of the $\Delta_{\alpha\beta k}$
brought about by the principal parts of the integrals have been neglected.

\subsubsection{Rotating-wave approximation and solution in the Floquet
representation}
\label{sect:RWA}
In a rotating-wave approximation (RWA), it is assumed that phase factors
$\exp[{\rm i}(\Delta_{\alpha\beta k}-\Delta_{\alpha'\beta'k'})t]$, with
$(\alpha,\beta, k)\neq(\alpha',\beta',k')$ in Eq.~(\ref{Improved
MasterEquation}) oscillate faster than all other time dependences and hence
can be neglected.  This argument applies, however, only to quasienergy spectra
without systematic degeneracies or quasidegeneracies. Indeed, the harmonic
potential we are presently dealing with has the peculiarity of equidistant
(quasi-) energy levels, cf.\ Eq.(\ref{SchrodingerFloquetSpectrum}), so that
additional terms have to be kept.  Here, the condition
$(\alpha-\beta,k)=(\alpha'-\beta',k')$ is sufficient to ensure
$\Delta_{\alpha\beta k}=\Delta_{\alpha'\beta'k'}$.  Therefore these terms have
to be kept in RWA.

Making the RWA, substituting
Eq.~(\ref{XFloquetFourier}) in Eq.~(\ref{Improved MasterEquation}), and
assuming an Ohmic bath as above, we obtain the time-independent master equation
\begin{eqnarray}
\nonumber
\dot\rho_{\alpha\beta} &=&
\frac{\gamma}{2}\left\{
(N+1)  \left(2\sqrt{(\alpha+1)(\beta+1)}\rho_{\alpha+1,\beta+1}
	-(\alpha+\beta)\rho_{\alpha\beta}\right) \right. \\
&&
\qquad +N\left.\left(2\sqrt{\alpha\beta}\rho_{\alpha-1,\beta-1}
	-(\alpha+\beta+2)\rho_{\alpha\beta}\right) \right\}.
\label{MasterEquationRWA}
\end{eqnarray}
The effective thermal-bath occupation number
\begin{equation}
N = \sum_{k} \left(c_k^{0}\right)^2(\mu^{0}+k\Omega)
n_{\rm th}(\mu^{0}+k\Omega)
\end{equation}
reduces to $N=n_{\rm th}(\omega_0)$ in the undriven limit.

Formally, this master equation coincides with that for the undriven
dissipative harmonic oscillator in rotating-wave approximation
\cite{Louisell73}. It has the stationary solution
\begin{equation}
\rho^{\rm as}_{\alpha\beta} =
\frac{1}{N+1}\left(\frac{N}{N+1}\right)^\alpha \delta_{\alpha\beta}.
\end{equation}

The density operator of the asymptotic solution is diagonal in this
representation and reads
\begin{equation}
\label{RhoStationary}
\rho_{\rm as}(t) = \sum_{\alpha=0}^\infty \rho^{\rm as}_{\alpha\alpha}
|\psi_\alpha(t)\rangle\langle\psi_\alpha(t)|.
\end{equation}

The basis $\left\{|\psi_\alpha(t)\rangle\right\}$ corresponds to the
``generalized Floquet states'' introduced in Ref.~\cite{GrahamHubner94}, i.e.,
they are centered on the classical asymptotic solution and diagonalize the
asymptotic density operator.

To get the variances of (\ref{RhoStationary}), we switch to the Wigner
representation,
\begin{equation}
W_{\rm as}(x,p,t) =
\sum_{\alpha=0}^\infty \rho^{\rm as}_{\alpha\alpha} W_\alpha(x,p,t) ,
\end{equation}
where
\begin{eqnarray}
&& W_\alpha(x,p,t) =
   \frac{(-1)^\alpha}{\pi}{\rm e}^{-z^2}L_\alpha(2z^2), \\
&& z^2 = \frac{1}{\hbar}\left(m\dot\xi^{0}_1(t)\dot\xi^{0}_2(t) x^2 -
   \left(\dot\xi^{0}_1(t)\xi^{0}_2(t)+
	\xi^{0}_1(t)\dot\xi^{0}_2(t)\right)px +
   \xi^{0}_1(t)\xi^{0}_2(t) p^2/m \right),
\end{eqnarray}
is the Wigner function corresponding to $|\psi_\alpha(t)\rangle$
\cite{SchradeMankoSchleichGlauber95}, with the
Laguerre polynomial $L_\alpha$. Using the sum rule \cite{GradshteynRyzhik}
\begin{equation}
\sum_{\alpha=0}^\infty \kappa^\alpha L_\alpha(x)
= (1-\kappa)^{-1}\exp\left(\frac{x\kappa}{\kappa-1}\right) ,
\end{equation}
we obtain the asymptotic solution in Wigner representation as
\begin{equation}
\label{SchrodingerWigner}
W_{\rm as}(x,p,t) = \frac{1}{\pi(2N+1)}{\rm e}^{-z^2/(2N+1)} .
\end{equation}
It is a Gaussian with the variances
\begin{eqnarray}
\sigma_{xx}(t) &=& \frac{\hbar}{m}(N+1/2) \xi^{0}_1(t)\xi^{0}_2(t), \\
\sigma_{xp}(t) &=& \hbar(N+1/2) \left(\dot\xi^{0}_1(t)\xi^{0}_2(t)
	+\xi^{0}_1(t)\dot\xi^{0}_2(t)\right)/2, \\
\sigma_{pp}(t) &=& \hbar m(N+1/2) \dot\xi^{0}_1(t)\dot\xi^{0}_2(t) .
\end{eqnarray}

To enable a comparison between the different equations of motions for the 
dissipative quantum system, we give for the master equation in RWA
(\ref{MasterEquationRWA}) also the corresponding partial differential
equation in Wigner representation.
For a derivation, we use the properties (\ref{Annihilation}) and
(\ref{Creation}) of the operators $\hat\Gamma$ and $\hat\Gamma^\dagger$, to get 
from the master equation (\ref{MasterEquationRWA}) for the density matrix
elements $\rho_{\alpha\beta}$ the corresponding operator equation
\begin{eqnarray}
\nonumber
\dot\rho &=& -\frac{\rm i}{\hbar}\Big[\hat H_{\rm S}(t),\rho \Big] \\
&& +\frac{\gamma}{2}\left\{
(N+1)\left( 2\hat\Gamma\rho\hat\Gamma^\dagger - \hat\Gamma^\dagger\hat\Gamma\rho
- \rho\hat\Gamma^\dagger\hat\Gamma \right)
+N\left( 2\hat\Gamma^\dagger\rho\hat\Gamma - \hat\Gamma\hat\Gamma^\dagger\rho -
\rho\hat\Gamma\hat\Gamma^\dagger\right)
\right\}.
\label{MasterEquationRWAoperator}
\end{eqnarray}
The dissipative part of this equation is the same as for the undriven
dissipative harmonic oscillator \cite{Louisell73}, but with the shift operators
for Floquet states instead of the usual creation and annihilation
operators.

Interestingly, the master equation in (\ref{MasterEquationRWAoperator})
now exhibits Lindblad form \cite{AlickiLendi87,Lindblad76}.

By substituting (\ref{ShiftOperator}), we get an operator equation which only
consists of position and momentum operators. Transforming them into the Wigner
representation, we find
\begin{equation}
\label{FokkerPlanckRWA}
L(t) = -\frac{1}{m}p\partial_x
+ \frac{\gamma}{2}( \partial_x x + \partial_p p) + k(t)x\partial_x
+\frac{\gamma}{2}\left( D_{xx}(t)\partial_x^2 +
 D_{xp}(t)\partial_x\partial_p + D_{pp}(t)\partial_p^2 \right)
\end{equation}
with the coefficients
\begin{eqnarray}
D_{xx}(t) &=& \hbar \xi^{0}_1(t)\xi^{0}_2(t) (N+1/2)/m ,
\label{DxxRWA} \\
D_{xp}(t) &=& \hbar \left(\dot\xi^{0}_1(t)\xi^{0}_2(t)
	+ \xi^{0}_1(t)\dot\xi^{0}_2(t)\right)(N+1/2)  ,
\label{DxpRWA} \\
D_{pp}(t) &=& m\hbar\dot\xi^{0}_1(t)\dot\xi^{0}_2(t)(N+1/2)  .
\label{DppRWA}
\end{eqnarray}

The fact that there are also dissipative terms in Eq.~(\ref{FokkerPlanckRWA})
containing derivatives with respect to $x$ is a consequence of the RWA:
Its effect is equivalent to using instead of (\ref{HamiltonianSB})
the coupling Hamiltonian
$H_{\rm SB}^{\rm RWA}=\sum_\nu g_\nu(ab_\nu^\dagger + a^\dagger b_\nu)$,
where $a$ and $b_\nu$ are the usual annihilation operators of the system
and the bath, respectively.
This introduces an additional coupling term $\propto p p_\nu$.
In the next subsection we show how to avoid this RWA, by going back to the
original Markov approximation, Eq.~(\ref{MasterEquation}).

\subsubsection{Fokker-Planck equation without rotating-wave approximation}
\label{sect:FPEwithoutRWA}
In the present case of a bilinear system, driven or not, for which the
classical motion is integrable, the knowledge of the classical dynamics
opens a more direct access also to the quantal time evolution. Specifically,
the interaction-picture position operator $\tilde x(t,t')$ for the
corresponding undamped quantum system is given by the solution of the
classical equation of motion in the limit $\gamma\to 0$, indicated
by the superscript ${}^{0}$.  In our case the classical solution is given by
(\ref{ClassicalSolution}). The corresponding interaction-picture position
operator reads
\begin{equation}
\tilde x(t,t') = -\hat x\frac{\partial G^{0}(t,t')}{\partial t'}
+ \frac{\hat p}{m} G^{0}(t,t') .
\label{WWpositionOperator}
\end{equation}
Inserting it into (\ref{MasterEquation}), we obtain a master equation in
Markov approximation with respect to the quasienergy spectrum without
expanding into Floquet states of the Schr\"odinger equation. Even with the
rotating-wave approximation avoided, the resulting equation has already a
simple structure: It is of the same form as the master equation derived in
Sect.~\ref{sect:MasterEq}, but with time-dependent transport coefficients
\begin{eqnarray}
\bar\gamma(t) &=&
	2\gamma\int_0^\infty{\rm d}\omega\,\omega
	\int_0^\infty {\rm d}\tau\, \sin(\omega\tau) G^{0}(t-\tau,t) , \\
D_{pp}(t) &=&
	-\frac{m\hbar}{\pi}\int_0^\infty{\rm d}\omega\,\omega
	\coth\left(\frac{\hbar\omega}{2k_{\rm B}T}\right)
	\int_0^\infty {\rm d}\tau\, \cos(\omega\tau)
        \left. \frac{\partial G^{0}(t-\tau,t')}{\partial t'}\right|_{t'=t},\\
\label{DxpImproved}
D_{xp}(t) &=&
	\frac{\hbar}{\pi}\int_0^\infty{\rm d}\omega\,\omega
	\coth\left(\frac{\hbar\omega}{2k_{\rm B}T}\right)
        \int_0^\infty{\rm d}\tau\, \cos(\omega\tau) G^{0}(t-\tau,t).
\end{eqnarray}
To evaluate these expressions, we substitute the undamped limit of
Eq.~(\ref{ClassicalGreenFunctionFourier}),
\begin{equation}
G^{0}(t,t') =
  \sum_{n,n'} c_n^{0}c_{n'}^{0}
  \sin\left[ \mu^{0}(t-t')+\Omega(nt-n't') \right],
\end{equation}
and exploit the sum rule (\ref{SumRule}) for the $c_n$, to find, as
in Sect.~\ref{sect:MasterEq},
\begin{equation}
\bar\gamma(t) = \gamma .
\end{equation}
The explicit time dependence in $G(t,t')$ results in a time dependence of the
coefficients $D_{pp}$ and $D_{xp}$. Averaging the transport coefficients 
over a period of driving, we
find for $D_{xp}$ with the sum rule (\ref{SumRule}) again the expression
(\ref{DxpHighCutoff}), as in Sect.~\ref{sect:MasterEq}.  Here, we have to
choose the cutoff $\omega_{\rm D}$ much larger than the relevant frequencies
$\mu^{0}+n\Omega$.

For $D_{pp}$ we find in an average over a period of driving
\begin{equation}
\label{DppFloquet}
D_{pp} = \frac{1}{2}m\hbar\sum_{n=-\infty}^\infty
	\left[ c_n^{0} (\mu^{0}+n\Omega) \right]^2
	\coth\frac{\hbar(\mu^{0}+n\Omega)}{2k_{\rm B}T}.
\end{equation}
Unlike the corresponding expression in the Sect.~\ref{sect:MasterEq},
Eq.~(\ref{DppSimple}), the diffusion $D_{pp}$ now accounts explicitly for the
quasienergies $\hbar(\mu^{0}+n\Omega)$ instead of the energy $\hbar\omega_0$.
Thus the quasispectrum approach is reflected solely by a driving-induced
modification of the momentum diffusion $D_{pp}$.

The Fokker-Planck equation for $W(x,p,t)$ is now of the same structure as
in the case of Markov approximation with respect to the unperturbed spectrum.
Therefore the solution and the Floquet-Wigner functions remain the
same, up to a different momentum diffusion $D_{pp}$.

In contrast to the Fokker-Planck equation with RWA in the last subsection, the
terms with $\partial_xx$ and $\partial_x^2$ are now absent. In addition, the
cross diffusion $D_{xp}$ in (\ref{DxpImproved}) is completely different, and
unrelated to the one in the RWA case (\ref{DxpRWA}).
It originates from a principal part that has been neglected in the
derivation of (\ref{FokkerPlanckRWA}).


\subsection{Asymptotics}
\subsubsection{The conservative limit}
In contrast to the Markov approximation with RWA in Sect.~\ref{sect:RWA}, the
variances in both Markov approximations without RWA still depend on the
friction $\gamma$.
To obtain the conservative limit $\gamma\to 0$ of these, we insert the Green
function (\ref{ClassicalGreenFunctionFourier}) into (\ref{SigmaXX}) and get
\begin{eqnarray}
\nonumber
\sigma_{xx}(t) &=&
-\frac{\gamma D_{pp}}{2m^2} \sum_{n,n'} c_n c_{n'}
\left(
f_1^2(t)\frac{{\rm e}^{\gamma t -
{\rm i}(2\mu+(n+n')\Omega)t}}{\gamma-{\rm i}(2\mu+(n+n')\Omega)}
\right.
\\ &&
-2f_1(t)f_2(t)\frac{{\rm e}^{\gamma t -
{\rm i}(n-n')\Omega t}}{\gamma-{\rm i}(n-n')\Omega}
\left.
+f_2^2(t)\frac{{\rm e}^{\gamma t +
{\rm i}(2\mu+(n+n')\Omega)t}}{\gamma+{\rm i}(2\mu+(n+n')\Omega)}
\right) .
\end{eqnarray}
In the limit of low damping, $\gamma \ll|\mu+n\Omega|$ for any integer $n$,
only the case $n=n'$ of the second term in the brackets remains.
Note that this condition is violated in parameter regions where the
Floquet index becomes an multiple of $\Omega$, as is the case along the
borderlines of the regions of stability in parameter space
(cf.\ Fig.~\ref{fig:stability}).

For the position variance, we get
\begin{equation}
\sigma_{xx}(t) = A\frac{D_{pp}}{m^2}\, \xi_1^{0}(t)\xi_2^{0}(t),
\end{equation}
where
\begin{equation}
A = \sum_{n=-\infty}^\infty \left(c^{0}_n\right)^2
\end{equation}
denotes a number of order unity.

In an analogous way, we find
\begin{eqnarray}
\sigma_{xp}(t) &=&
A\frac{D_{pp}}{2m}
\left(\dot \xi_1^{0}(t)\xi_2^{0}(t)+\xi_1^{0}(t)\dot\xi_2^{0}(t)\right),\\
\sigma_{pp}(t) &=& AD_{pp}\, \dot \xi_1^{0}(t) \dot \xi_2^{0}(t).
\end{eqnarray}
Besides the prefactor, these variances are the same as for the master equation
with RWA in Sect.~\ref{sect:FPEwithoutRWA}.

Moreover, in this limit $\gamma\to 0$, all diagonal elements $W_{nn}(x,p,t)$ 
are Floquet functions with the quasienergies $\mu_{nn}=0$.
However, they are different from the Wigner representation of the
stationary solutions (\ref{SchrodingerWigner}) of the corresponding
Schr{\"o}dinger equation, which are of course
solutions of (\ref{FokkerPlanck}) with $\gamma=0$.
Due to the degeneracy of the Floquet indices, this is no contradiction.
The $W_{nn}(x,p,t)$ can be viewed as dissipation-adapted Floquet functions.

For consistency, we check the uncertainty relations for the asymptotic
solution. It is satisfied if the variances fulfill the inequality
\begin{equation}
\left| \begin{array}{cc}
   \sigma_{xx}(t) & \sigma_{xp}(t) \\
   \sigma_{xp}(t) & \sigma_{pp}(t)
\end{array}\right|
= \left(\frac{D_{pp}A}{m}\right)^2
\geq \hbar^2/4 ,
\end{equation}
which we have verified numerically for the case of the Mathieu oscillator.

\subsubsection{The high-temperature limit}

In the limit of high temperatures $k_{\rm B}T\gg\hbar\omega_{\rm D}$,
we expect the Fokker-Planck equation
for the Wigner function to give the Kramers equation for the classical
Brownian motion \cite{ZerbeJungHanggi94}, i.e.~an equation of the form
(\ref{FokkerPlanck}) with the diffusion constants $D_{xp}=0$ and
$D_{pp}=mk_{\rm B}T$.

In the standard approach (Sect.~\ref{sect:SimpleMarkov}) and the quasispectrum
approach without RWA (Sect.~\ref{sect:FPEwithoutRWA}), the Fokker-Planck
equation is already of the required structure.  With $\psi(1)=C$
\cite{GradshteynRyzhik} the cross diffusion $D_{xp}$ vanishes in the
high-temperature limit. For $D_{pp}$, we use $\coth x=1/x + {\cal O}(x)$ and
get
\begin{equation}
D_{pp}=mk_{\rm B}T\sum_n \left(c_n^{0}\right)^2 (\mu^{0}+n\Omega).
\end{equation}
With the sum rule (\ref{SumRule}), this reduces to $D_{pp}=mk_{\rm B}T$.

In the quasispectrum approach with RWA in Sect.~\ref{sect:RWA},
the variances and diffusion constants scale with $N+1/2$.
This factor reads, in the high-temperature limit,
\begin{equation}
N+\frac{1}{2} = \sum_n \left(c_n^{0}\right)^2\frac{k_{\rm B}T}{\hbar}
=A\frac{k_{\rm B}T}{\hbar}.
\end{equation}
Therefore the diffusion constants $D_{xx}$ and $D_{xp}$ remain finite
and the Fokker-Planck operator (\ref{FokkerPlanckRWA}) does not
approach the Kramers limit for high temperatures.
Nevertheless the asymptotic variances in RWA coincide for high temperatures,
with the classical result in the limit $\gamma\to 0$.

\section{Numerical results}
\label{sect:NumericalResults}
In this section, we compare our approximate results to exact ones, obtained
from the path-integral solution in Ref.~\cite{ZerbeHanggi95}.  Specifically,
we give the numerical results for the Mathieu oscillator, i.e., we use
\begin{equation}
\label{Mathieu}
k(t) = m\left( \omega_0^2 + \varepsilon\cos\Omega t \right).
\end{equation}
This is an experimentally important case in view of the fact that it
describes the Paul trap \cite{Paul90}.

By inserting (\ref{Mathieu}) and the ansatz (\ref{FourierAnsatz})
into (\ref{ClassicalY}), we obtain the tridiagonal recurrence relation
\begin{equation}
\varepsilon c_{n-1}
+ 2\left(\omega_0^2-\gamma^2/4-(\mu + n\Omega)^2 \right) c_n
+\varepsilon c_{n+1} = 0.
\end{equation}
From this equation, the classical Floquet index $\mu$ and the
Fourier coefficients $c_n$ are determined numerically by continued fractions
\cite{Risken84}.

In the figures we use the scaled quantities $\bar t=\Omega t/2$,
$\bar \omega_0=2\omega_0/\Omega$ and $\bar\varepsilon=2\varepsilon/\Omega^2$.
The external period thus takes the value $\bar T=\pi$.
Position and momentum are scaled via $\bar x=(2\hbar/m\Omega)^{-1/2}x$ and
$\bar p=(m\hbar\Omega/2)^{-1/2}p$, respectively.
The overbar for the scaled quantities has been suppressed in the figures.

\begin{figure}
\centerline{ \psfig{width=10cm,figure=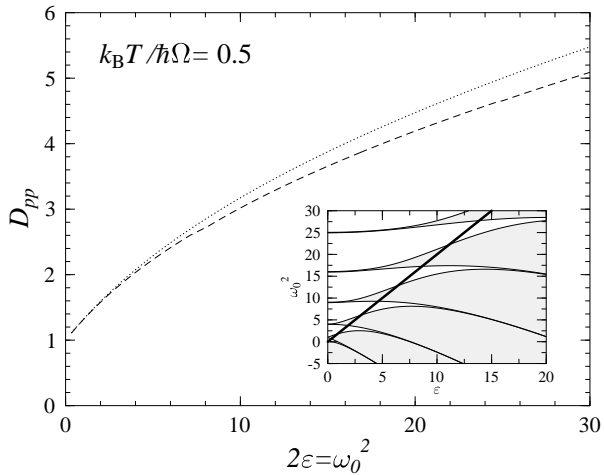} }
\caption{The diffusion constant $D_{pp}$ for the simple (dotted) and the 
improved (dashed) Markov approximation compared to the time-average of the
exact value in units of the classical
diffusion constant $mk_{\rm B}T$ for $k_{\rm B}T=0.5\,\hbar\Omega$.
The parameters $\omega_0^2$ and $\varepsilon$ are indicated by the full line
in the insert (units as in Fig.~\protect\ref{fig:stability}).}
\label{fig:diffusion}
\end{figure}

The influence of the quasienergies on the equation of motion
(\ref{FokkerPlanck}) is given by different diffusion coefficients $D_{pp}$.
In Fig.~\ref{fig:diffusion}, we compare the momentum-diffusion coefficients
between the Markov approximation with respect to the unperturbed spectrum,
given by Eq.~(\ref{DppSimple}), and the Markov approximation that relates
to the quasienergy spectrum, given by Eq.~(\ref{DppFloquet}).
We have scaled the values to the classical momentum-diffusion coefficient
$mk_{\rm B}T$.
The parameters $\omega_0^2$ and $\varepsilon$ are varied along the full
line in the inset.
Note that within the unstable regimes, perturbation theory is not valid.
Nevertheless, Eq.~(\ref{DppFloquet}) gives a smooth interpolation.
The discrepances become most significant for strong driving and large
$\omega_0^2$.
For both, low driving amplitude $\varepsilon\ll\omega_0^2$ and
high temperature $T\gg\hbar\omega_0/k_{\rm B}$, the difference 
vanishes. 

\begin{figure}
\centerline{ \psfig{width=10cm,figure=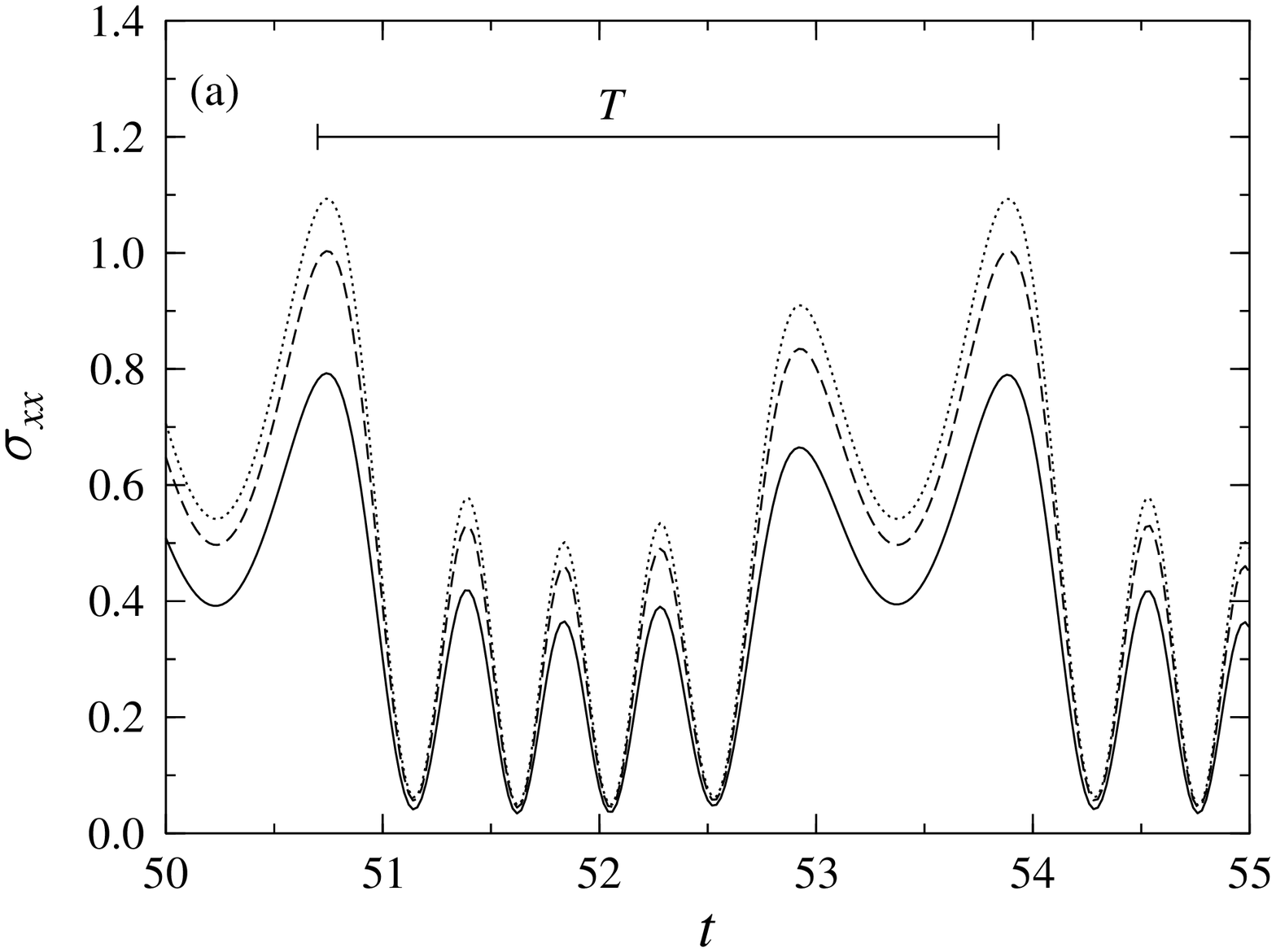} }
\centerline{ \psfig{width=10cm,figure=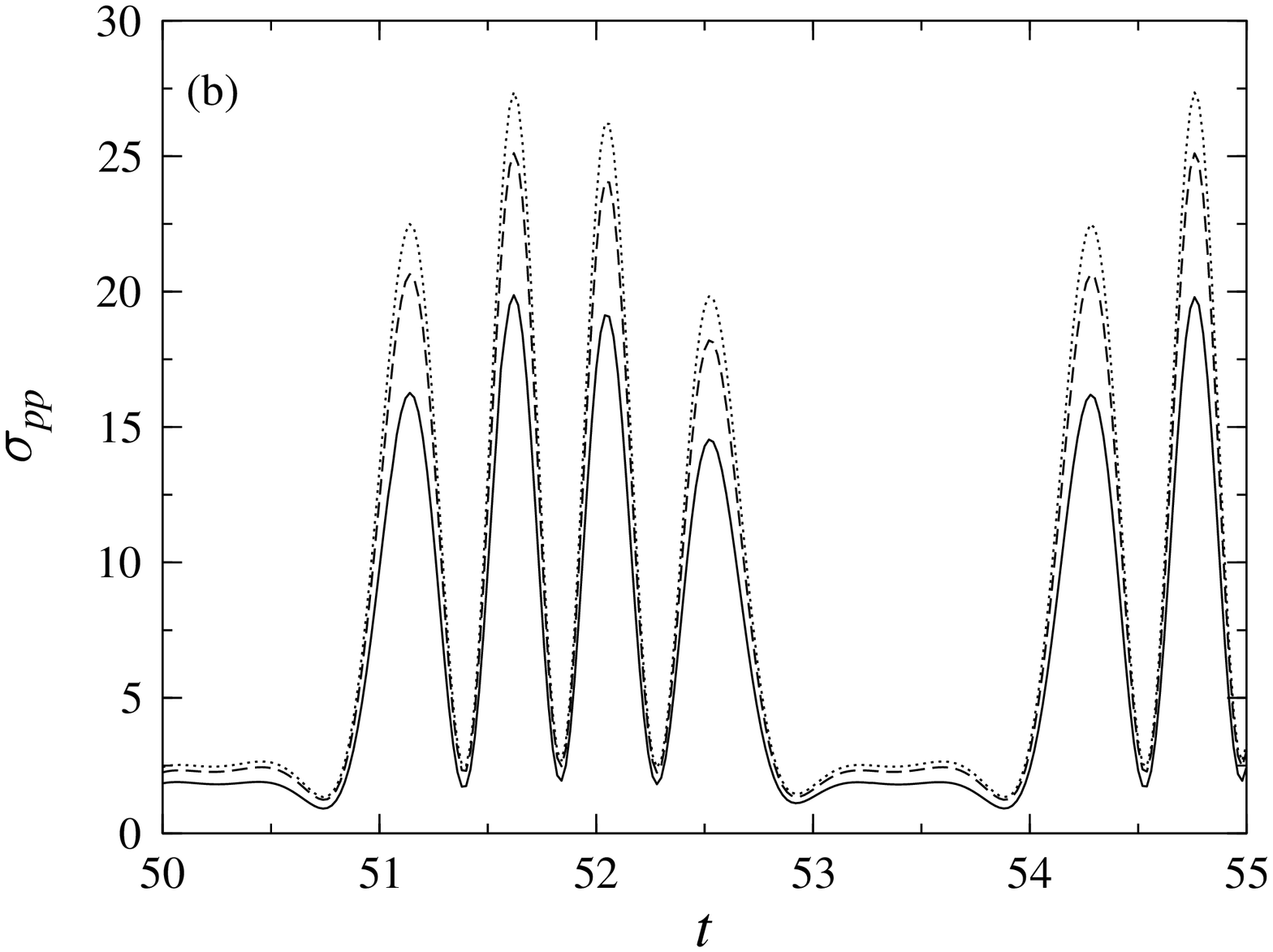} }
\caption{The asymptotic variances $\sigma_{xx}(t)$ (a) and $\sigma_{pp}(t)$ (b)
with period $T=2\pi/\Omega$ for the simple (dotted) and the
improved (dashed) Markov approximation, compared to the exact result
(full line) for the parameters $\varepsilon=7\,\Omega^2$,
$\omega_0^2=6.5\,\Omega^2$, $k_{\rm B}T=0.5\,\hbar\Omega$ and
$\gamma=\Omega/20$.
The scaled driving period $\bar T=\pi$ is indicated in panel (a).}
\label{fig:variance:FPE}
\end{figure}

The variances $\sigma_{xx}(t)$ and $\sigma_{pp}(t)$
of the Markov approximations without RWA are compared
against the exact results \cite{ZerbeHanggi95} in the panels~%
\ref{fig:variance:FPE}a and \ref{fig:variance:FPE}b.
The chosen driving parameters $\omega^2=6.5\,\Omega^2$ and
$\varepsilon=7\,\Omega^2$ lie inside the fifth stable zone
($\mu=4.53513\,\Omega/2$). The temperature $k_{\rm B}T=0.5\,\hbar\Omega$
is sufficiently large, but with quantum effects still appreciable.
We note that the improved Markov treatment in Sect.~\ref{sect:FPEwithoutRWA},
that accounts for the quasienergy differences, agrees better
with the exact prediction.
In the Figure we depict asymptotic times $\bar t>50$, where transient
effects have already decayed. The asymptotic co-variance elements retain the
periodicity $\bar T=\pi$ of the external driving.
For the chosen parameters the relative error is reduced by the use of the
improved Markov scheme by approximately 30\%.

\begin{figure}
\centerline{ \psfig{width=10cm,figure=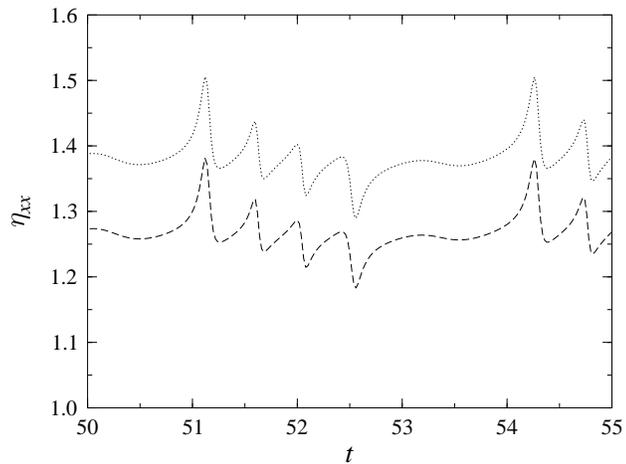} }
\caption{Relative error
$\eta_{xx}(t)=\sigma_{xx}^{\rm Markov}(t)/\sigma_{xx}^{\rm exact}(t)$
for the position variances of panel~\protect\ref{fig:variance:FPE}a.}
\label{fig:RelativeError}
\end{figure}

The relative error
$\eta_{xx}(t)=\sigma_{xx}^{\rm Markov}(t)/\sigma_{xx}^{\rm exact}(t)$
of the position variance for these two Markov approximations is depicted in
panel~\ref{fig:RelativeError}.
Note that the maximal deviations do not occur in the extrema, but happen to
occur in the regions with negative slope.

\begin{figure}
\centerline{ \psfig{width=10cm,figure=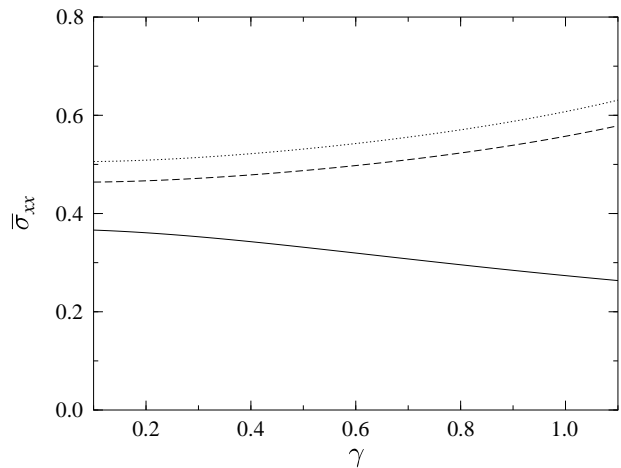} }
\caption{The time averaged variance $\bar\sigma_{xx}(t)$ for the simple (dotted)
and the improved (dashed) Markov approximation, compared to the exact result
(full line) for the parameters $\varepsilon=7\,\Omega^2$,
$\omega_0^2=6.5\,\Omega^2$ and $k_{\rm B}T=0.5\,\hbar\Omega$.}
\label{fig:variance:average}
\end{figure}

As depicted with Fig.~\ref{fig:variance:average}, the quality of both
Markov approximations worsens with increasing dissipation strength $\gamma$.
This reflects the breakdown of the weak coupling approach when strong
friction is ruling the system dynamics.

\begin{figure}
\centerline{ \psfig{width=10cm,figure=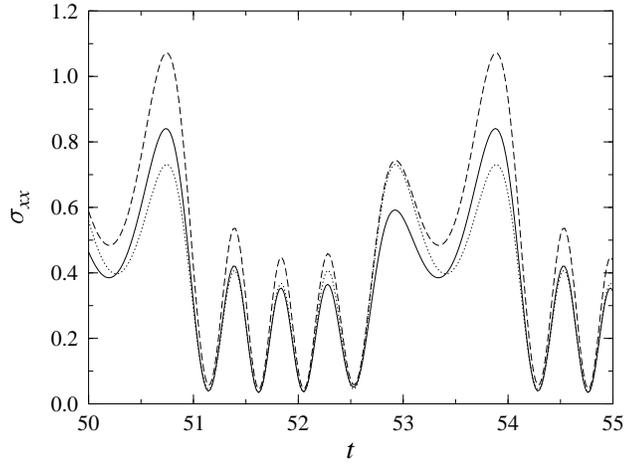} }
\caption{Position variances obtained with the Markov approximation with
respect to the quasienergy spectrum with (dotted) and without (dashed) RWA,
compared to the exact result (full line) for
$\gamma=\Omega/10$ for $k_{\rm B}T=0.5\,\hbar\Omega$.
The driving parameters are $\varepsilon=7\,\Omega^2$ and
$\omega_0^2=6.5\,\Omega^2$.}
\label{fig:variance:rwa}
\end{figure}

Results for the Markovian treatment within RWA, given in Sect.~\ref{sect:RWA},
are depicted for the position variance $\sigma_{xx}(t)$ in
Fig.~\ref{fig:variance:rwa}. The driving parameters are the same as in
Fig.~\ref{fig:variance:FPE}.
For this example, the quality of agreement to the exact result is similar for
both Markov approximations.
Nevertheless, the solution without RWA yields---up to a scale---a better
overall agreement with the exact behaviour over a full driving period $T$.

%
\section{Conclusion}

We have used the parametrically driven harmonic oscillator as a simple
working example to compare various versions of the Markovian approach to the
quantum dynamics of periodically driven systems with dissipation, and to
provide a synopsis of a number of alternative representations,
each of which emphasizes different aspects of the same underlying physics.

The principal distinction to be made among possible Markovian approaches to
the driven dissipative dynamics, refers to the degree to which changes in
dynamical and spectral properties of the central system due to the driving are
taken into account. In the crudest treatment, the non-unitary terms in the
master equation are derived ignoring the explicit time dependence of the
Hamiltonian, and the driving appears only in the unitary term. An improved
master equation is obtained if the central system and the driving are coupled
to the heat bath as one whole. The energy-domain quantity relevant for all
subsequent developments is then the quasienergy spectrum, obtained within the
Floquet formalism, instead of the unperturbed spectrum. In the time domain,
the quantities entering the dissipative terms of the master equation, such as
Heisenberg-picture operators of the central system, gain an explicit time
dependence with the periodicity of the driving. As a bonus, the Floquet
treatment of the central system with driving yields a well-adapted basis, the
set of eigenstates of the Floquet operator. Representing the master equation
in this basis completely removes the unitary term.

Besides the differences in representation, the use of the improved
Floquet-Markov approximation in Sect.~\ref{sect:FPEwithoutRWA} results mainly
in a modified momentum diffusion that dependss on the quasienergy spectrum
instead of the unperturbed spectrum of the central system.  The difference
becomes significant in the limits of strong driving amplitude and low
temperature.  An {\it additive\/} time-dependent external force, applied in
addition to or instead of the parametric driving, undergoes a renormalization
which vanishes, however, in the case of an Ohmic bath.

Even within the improved Markov approach, finer levels of approximation can be
distinguished. A significant simplification of the master equation is achieved
by a rotating-wave approximation, i.e.\ here, by neglecting reservoir-induced
virtual transitions between {\it Floquet\/} states of the central system that
violate {\it quasienergy\/} conservation. The resulting master equation has
Lindblad form, with creation and annihilation operators acting on Floquet
states, and thus manifestly generates a dynamical semigroup. This is not the
case if the RWA is avoided. Apparently a drawback, the lack of a Lindblad
structure in the master equation without RWA faithfully reflects the failure
of the Markov approximation on short time scales.

An analogous situation as with the Lindblad form of the master equation arises
with its Floquet structure. If all coefficients are at most periodically time
dependent, then the equation of motion for the reduced density operator
complies with the conditions for applicability of the Floquet theorem. As a
consequence, the solutions can be cast in Floquet form, i.e., can be written
as eigenfunctions of a generalized non-unitary Floquet operator that generates
the evolution of the density operator over a single period. Since all variants
of the Markov approximation discussed herein truncate the memory of the
central system on time scales shorter than the period of the driving, the
corresponding master equations have Floquet structure throughout. The exact
path-integral solution, in contrast, allows for memory effects of unlimited
duration and thereby generally prevents the consistent definition of a
propagator over a single period.

Additional insight is gained by discussing the dynamics in terms of
phase-space distributions, specifically, in terms of the Wigner representation
of the density operator and its equation of motion.
In this representation, the Floquet formalism is a useful
device to construct and classify solutions. Since all Fokker-Planck equations
obtained are time periodic, as are the corresponding master equations, their
solutions may be written as eigenstates of a {\it Wigner-Floquet operator\/}
(the Fokker-Planck operator evolving the Wigner function, integrated over a
single period), or {\it Wigner-Floquet states\/} in short. They represent the
quasiprobability distributions closest to the Floquet solutions of the
corresponding classical Fokker-Planck equation.

Wigner-Floquet states with a purely real quasienergy correspond to asymptotic
solutions. They are not literally stationary but retain the periodic time
dependence of the driving. Since we are here dealing with a linear system, the
asymptotic quasiprobability distributions follow the corresponding classical
limit cycles. In the case of parametric driving, these limit cycles
are trivial and correspond to a fixed point at the origin. A time dependence
arises only by the periodic variation of the shape of the asymptotic
distributions.

Concluding from a numerical comparison of certain dynamical quantities, for
the specific case of the Mathieu oscillator, the attributes ``simple'' and
``improved'' for the two basic Markovian approaches prove adequate. Results
for the Markov approximation based on the quasienergy spectrum show
consistently better agreement with the exact path-integral solution than those
for the Markov approximation with respect to the unperturbed
spectrum. However, even in parameter regimes where the respective
approximations are expected to become problematic, the differences in quality
are not huge and the agreement with the exact solution is generally
good. Technical advantages of the Markov approximation in general and of its
various ramifications---easy analytical and numerical tractability, desirable
formal properties such as Floquet or Lindblad form of the master
equation---can justify to accept their quantitative inaccuracy.

%
\section*{Acknowledgments}
Financal support of this work by the Deutsche Forschungsgemeinschaft
(Grant No.~Di 511/2-1 and Ha 1517/14-1) is gratefully achnowledged.
We thank Christine Zerbe for providing us the numerical code for the path
integral solution and Gert-Ludwig Ingold for helpful discussions.
%
\appendix
\section{Solution of the Characteristic Equations}
\label{appendix:characteristic}
In this appendix, we solve the equation of motion for the Wigner
function by the method of characteristics.
For simplicity, we use here units with $m=1$.
We write $W(x,p,t)$ as
\begin{equation}
\label{FourierEikonal}
W(x,p,t) = \int {\rm d}X {\rm d}P\,
{\rm e}^{{\rm i}xX + {\rm i}pP} {\rm e}^{S(X,P,t)} .
\end{equation}
By this ansatz, equation (\ref{FokkerPlanck}) is transformed to the
quasilinear partial differential equation
\begin{equation}
\label{partDGL}
{\cal F}(X,S_X,P,S_P,t,S_t) = 0
\end{equation}
for $S(X,P,t)$, where ${\cal F}$ is given by
\begin{equation}
{\cal F}
= S_t - XS_P + \gamma PS_P + k(t)PS_X + \gamma D_{pp}P^2 + \gamma D_{xp}XP.
\end{equation}
We denote the partial derivatives of $S(X,P,t)$ with respect to $X$,
$P$ and $t$ by $S_X$, $S_P$ and $S_t$, respectively.

The characteristic equations \cite{Kamke79b} of (\ref{partDGL}) are given by
\begin{eqnarray}
\label{characteristict}
\dot t &=& \frac{\partial {\cal F}}{\partial S_t} = 1 ,\\
\label{characteristicX}
\dot X &=& \frac{\partial {\cal F}}{\partial S_X} = k(t) P ,\\
\label{characteristicP}
\dot P &=& \frac{\partial {\cal F}}{\partial S_P} = \gamma P - X ,\\
\label{characteristicSX}
\dot S_X &=& -\frac{\partial {\cal F}}{\partial X} = S_P - \gamma D_{xp}P ,\\
\label{characteristicSP}
\dot S_P &=& -\frac{\partial {\cal F}}{\partial P}
	= -\gamma S_P - k(t)S_X - 2\gamma D_{pp}P - \gamma D_{xp}X ,\\
\label{characteristicSt}
\dot S_t &=& -\frac{\partial {\cal F}}{\partial t}
	= -\frac{{\rm d}k(t)}{{\rm d}t} P S_X ,
\end{eqnarray}
whose solutions give the characteristics of the partial
differential equation (\ref{partDGL}).

Equation (\ref{characteristict}) means that the characteristics can be
parameterized by the time $t$.  Instead of equation (\ref{characteristicSt}),
we will use (\ref{partDGL}) to get an expression for $S_t$.  So we only have
to solve (\ref{characteristicX})--(\ref{characteristicSP}).  The solutions of
these equations can be traced back to the fundamental solutions $f_i(t)$ of the
classical equation of motion (\ref{ClassicalX}).

From (\ref{characteristicX}) and (\ref{characteristicP}), we find
\begin{equation}
\ddot P - \gamma\dot P + k(t)P = 0 .
\end{equation}
This is simply the classical equation of motion with a negative
damping constant. Therefore the solutions for $X$ and $P$ read
\begin{eqnarray}
\label{CharacteristicSolutionP}
P(t) &=& -c_{1+}{\rm e}^{\gamma t}f_2(t)
	+ c_{2+}{\rm e}^{\gamma t}f_1(t) , \\ 
\label{CharacteristicSolutionX}
X(t) &=& c_{1+}{\rm e}^{\gamma t}\dot f_2(t)
	- c_{2+}{\rm e}^{\gamma t}\dot f_1(t) ,
\end{eqnarray}
where  $c_{i+}$ denote integration constants.

From (\ref{characteristicSX}) and (\ref{characteristicSP}) we find for
$S_X$
\begin{equation}
\label{characteristicSX2}
\ddot S_X + \gamma\dot S_X + k(t)S_X = -2 \gamma DP,
\end{equation}
which is the classical equation of motion with an inhomogeneity.
The effective diffusion constant $D$ is given by
\begin{equation}
\label{Deffective}
D=D_{pp}+\gamma D_{xp}.
\end{equation}
With the integration constants $c_{i-}$, we integrate
(\ref{characteristicSX2}) with the Green function
(\ref{ClassicalGreenFunction}) to
\begin{equation}
S_X(t) =
c_{1-}f_1(t) + c_{2-}f_2(t) -2\gamma D\int_{t_0}^t {\rm d}t'\,G(t,t')P(t') ,
\end{equation}
and get by use of (\ref{characteristicSX})
\begin{equation}
S_P(t) = c_{1-}\dot f_1(t) + c_{2-}\dot f_2(t)
- 2\gamma D\int_{t_0}^t {\rm d}t'\, \frac{\partial G(t,t')}{\partial t}P(t')
+ \gamma D_{xp}P(t) .
\end{equation}
By inserting
\begin{equation}
P(t') = G(t,t')X(t) + \frac{\partial G(t,t')}{\partial t}P(t),
\end{equation}
obtained from Eqs.~(\ref{CharacteristicSolutionP}) and
(\ref{CharacteristicSolutionX}), we get a result for $S_X$ and $S_P$
that depends only on the endpoints of the characteristics.
Now together with Eq.~(\ref{partDGL}), we have an expression for
${\rm grad}\, S(X,P,t)=(S_X,S_P,S_t)$, which can be integrated to
\begin{eqnarray}
\nonumber
S(X,P,t) &=& \Big( c_{1-}f_1(t) + c_{2-}f_2(t) \Big) X
	   + \Big( c_{1-}\dot f_1(t) + c_{2-}\dot f_2(t) \Big)P \\
	 &&  -\frac{1}{2}\sigma_{xx}(t,t_0)X^2
	   -\sigma_{xp}(t,t_0)XP -\frac{1}{2}\sigma_{pp}(t,t_0)P^2 ,
\label{S(X,P,t)}
\end{eqnarray}
with
\begin{eqnarray}
\label{SigmaXX_t0}
\sigma_{xx}(t,t_0) &=&
	2\gamma D\int_{t_0}^t {\rm d}t' \left[ G(t,t') \right]^2 , \\
\label{SigmaXP_t0}
\sigma_{xp}(t,t_0) &=&
	2\gamma D\int_{t_0}^t {\rm d}t'
	G(t,t') \frac{\partial}{{\partial}t}G(t,t') , \\
\label{SigmaPP_t0}
\sigma_{pp}(t,t_0) &=& -\gamma D_{xp}
	+ 2\gamma D\int_{t_0}^t {\rm d}t'
	\left[ \frac{\partial}{{\partial}t}G(t,t')\right]^2 .
\end{eqnarray}
By inserting $S(X,P,t)$ into (\ref{FourierEikonal}), we find a
solution for the Wigner function $W(x,p,t)$.

The integration constants $c_{i\pm}$ are of course constant along the 
characteristics.
Therefore the
Poisson brackets between the expressions $c_{i\pm}(X,S_X,P,S_P,t)$ and ${\cal F}(X,S_X,P,S_P,t,S_t)$ vanish \cite{Kamke79b}.
By transforming back from Fourier space to real space, one finds
that the operators
$\hat c_{i\pm} \equiv
 c_{i\pm}(-{\rm i}\partial_x,-{\rm i}x,-{\rm i}\partial_p,-{\rm i}p,t)$
commute with the operator $\partial_t-L(t)$, whose nullspace is the solution
of the equation of motion.
Therefore, the
$\hat c_{i\pm}$ are shift operators in the subspace of solutions, i.e.~if
$W(x,p,t)$ is a solution of (\ref{FokkerPlanck}), then $\hat c_{i\pm}W(x,p,t)$
is also a solution.

For the $\hat c_{i\pm}$ we find
\begin{eqnarray}
\hat c_{1+} &=&
\frac{1}{2} \left( f_1(t)\partial_x + \dot f_1(t)\partial_p \right), \\
\hat c_{2+} &=&
\frac{1}{2} \left( f_2(t)\partial_x + \dot f_2(t)\partial_p \right) \\
\nonumber
\hat c_{1-} &=&
{\rm i}\dot f_2(t)
\Big( x+\sigma_{xx}(t,t_0)\partial_x+\sigma_{xp}(t,t_0)\partial_p\Big) \\
&& -{\rm i} f_2(t)
\Big( p+\sigma_{xp}(t,t_0)\partial_x+\sigma_{pp}(t,t_0)\partial_p \Big) \\
\nonumber
\hat c_{2-} &=&
-{\rm i}\dot f_1(t)
\Big( x+\sigma_{xx}(t,t_0)\partial_x+\sigma_{xp}(t,t_0)\partial_p\Big) \\
&& +{\rm i} f_1(t)
\Big( p+\sigma_{xp}(t,t_0)\partial_x+\sigma_{pp}(t,t_0)\partial_p \Big) .
\end{eqnarray}

Note that because of the linear structure of the characteristic equations,
there is no ambiguity concerning the ordering of operators.

The operators $Q_{i+}(t)$, used above, are proportional to the $\hat c_{i+}$.
%
\section{The additively driven harmonic oscillator}
\label{appendix:LinearForce}

In this appendix we present the Markovian master equation within the
quasi-spectrum approach when the parametric oscillator is subjected to
additional additive driving $-\hat x F(t)$, i.e.
\begin{equation}
\hat {\cal H}(t) = \hat H_{\rm S}(t) - \hat x F(t).
\label{HamiltonainXF}
\end{equation}
With $\hat H_{\rm S}(t)$ being a time-independent harmonic oscillator,
i.e., $k(t)=m\omega_0^2$,
the corresponding Markovian master equation in RWA for the dissipative system
has already been given in \cite{GrahamHubner94}.
Herein we generalize these results for the combined time-dependent
system Hamiltonian in (\ref{HamiltonainXF}).

It is known that the only effect of the driving force $F(t)$ on the
(quasi-) energy spectrum of a parametrically driven harmonic oscillator
is an overall level shift \cite{PopovPerelomov70}.
Thus the level separations remain unaffected and we expect no change
in the dissipative part of the master equation (\ref{MasterEquationSimple}).

The classical equation of motion, which is also obeyed by the
interaction-picture position operator, now reads
\begin{equation}
m\ddot x + k(t)x = F(t),
\end{equation}
and can be integrated to yield the interaction-picture position operator
\begin{equation}
\tilde x(t,t') = - \hat x\frac{\partial G^{0}(t,t')}{\partial t'}
+ \frac{\hat p}{m} G^{0}(t,t') 
+\frac{1}{m}\int_{t'}^t {\rm d}t''\,G^{0}(t,t'')F(t'').
\label{WWpositionOperatorF}
\end{equation}
Thus we obtain a c-number correction to the interaction-picture position
operator (\ref{WWpositionOperator}), given by the third term.
After inserting (\ref{WWpositionOperatorF}) into (\ref{MasterEquation}), the
generalized Markov approximation emerges as
\begin{eqnarray}
\dot\rho_{\rm S} &=&
   \ldots+ \frac{\rm i}{\hbar}F(t)\big[\hat x,\rho_{\rm S}\big]\\
&& -\frac{\rm i}{\hbar^2}\sum_\nu g_\nu^2 \int_0^\infty {\rm d}\tau\,
   A_\nu(\tau) \big[\hat x,\rho_{\rm S}\big]\frac{2}{m}\int_t^{t-\tau} {\rm d}t'
   G^{0}(t-\tau,t')F(t') .
\end{eqnarray}
The dots denote the old result for $F(t)=0$, given by the right hand side
of Eq.~(\ref{MasterEquationSimple}).
The term in the first line stems from the reversible part of the master
equation (\ref{MasterEquation}); the second one is a correction of the
driving force due to the interaction with the bath.
Thus the equation of motion for the density operator has the structure
\begin{equation}
\dot\rho_{\rm S} = \ldots + 
\frac{\rm i}{\hbar} \tilde F(t)\big[\hat x,\rho_{\rm S}\big]
\label{MasterEquationF}
\end{equation}
with an effective total driving force
\begin{equation}
\tilde F(t) = F(t)+\frac{2}{m\pi}\int_0^\infty {\rm d}\omega\,I(\omega)
\int_0^\infty {\rm d}\tau\,\sin\omega\tau \int_t^{t-\tau} {\rm d}t'
G^{0}(t-\tau,t') F(t') .
\label{Feffective}
\end{equation}

Note that the dissipative parts of (\ref{MasterEquationF}) are not
affected by the additive driving force $F(t)$.
This makes explicit, that we must use a parametric time-dependence to 
study differences in the dissipative parts resulting from the Markov
approximation with respect to the energy spectrum versus the Markov
approximation with respect to the quasienergy spectrum.

With an Ohmic bath, $I(\omega)=m\gamma\omega$, the integral in
(\ref{Feffective}) vanishes and we obtain $\tilde F(t)=F(t)$.  Thus in
contrast to an explicit parametric time dependence $k(t)$ in the quadratic
part of the Hamiltonian, the time dependence of an additive force, in this
case, does not change the Markovian master equation of the dissipative system.


%
%
\end{document}